\documentclass[journal]{IEEEtran}

\ifCLASSINFOpdf
\else
\fi
\IEEEoverridecommandlockouts
\usepackage{cite}
\usepackage{amsmath,amssymb,amsfonts}
\usepackage{algorithmic}
\usepackage{graphicx}
\usepackage{textcomp}
\usepackage{xcolor}
\usepackage{graphicx}
\usepackage[ruled,vlined]{algorithm2e}
\usepackage{flushend} 
\usepackage{booktabs}

\graphicspath{ {./images/} }
\def\BibTeX{{\rm B\kern-.05em{\sc i\kern-.025em b}\kern-.08em T\kern-.1667em\lower.7ex\hbox{E}\kern-.125emX}}
\begin{document}

\abovedisplayskip=-2pt 
\abovedisplayshortskip=-7pt 
\belowdisplayskip=5pt
\belowdisplayshortskip=5pt
\setlength\abovecaptionskip{1pt}
\setlength{\textfloatsep}{1pt plus 1.0pt minus 2.0pt}
\setlength{\floatsep}{1pt plus 1.0pt minus 2.0pt}


\hyphenation{op-tical net-works semi-conduc-tor}

%

\title{A Transactive Energy Market Framework Considering Network Constraints and Fairness}

%
%
%

        
\author{Fredmar~Asarias,~\IEEEmembership{Member,~IEEE,}
        and~Michael~Angelo~Pedrasa,~\IEEEmembership{Member,~IEEE}


\thanks{The authors are with the Electrical and Electronics Engineering Institute, University of the Philippines Diliman, Quezon City, PH. Emails: \{fnasarias,mapedrasa\}@up.edu.ph}

\thanks{F. Asarias is also with the Department of Science and Technology - Advanced Science and Technology Institute, Quezon City, PH.}
}

\maketitle

\begin{abstract}
The continuous penetration of distributed energy resources (DER) in the electric power grid is driving a new paradigm shift towards transactive energy system (TES), an active and more sustainable system characterized by distributed generation and energy exchanges among consumers and producers in the network. This transition, however, comes with challenges such as dealing with the nonlinear and non-convex power flows of the system, determining an optimal transaction price to maximize overall system welfare, and ensuring fairness for all participants. In this paper, we propose a three-stage  transactive energy framework that aims to address these challenges. In the first stage, the cost without trading is calculated which will serve as the reference in the profit maximization problem in the next stage. DER dispatch, power flows and initial transaction payments/incentives of the participants will then be determined in the second stage. A benefit allocation algorithm is applied in the third control stage to determine the optimal transaction price and final payments/incentives that will ensure fairness for trading participants. The proposed framework was tested in an IEEE 33-bus system and results show that fair benefits are given for all participants during trading and the system operates within the network and economic constraints.
\end{abstract}


%
\IEEEpeerreviewmaketitle

\section{Introduction}
%
%
%
%

 




\IEEEPARstart{F}{or} the longest time, the electrical power grid has been a vertically integrated system characterized by few companies controlling the generation, transmission and distribution of power and energy. This type of framework gives monopoly to large utilities and leaves customers adhering to the price dictated by these companies. But over the years, DERs such as rooftop solar, storage, and electric vehicles (EV) are being integrated to homes, commercial and industrial establishments that gives the capability for independent power generation and enables exportation of surplus power back to the grid for incentives from the utility. In the USA, DER installations are projected to exceed 60,000 MW capacity in 2024 \cite{federal2018distributed}. With this, the idea of being able to generate own power and manage it more effectively by selling/buying to/from other consumers/power producers has been explored in the recent years and this has paved the way to the emergence of \textit{transactive energy systems} that is defined as a "system of economic and control mechanisms that allows the dynamic balance of supply and demand across the entire electrical infrastructure using value as a key operational parameter" based from the Gridwise Architecture Council \cite{council2015gridwise}.

For TES, trading and scheduling mechanism can be considered as the most vital aspect of the system. This mechanism determines the optimal schedule and dispatch of respective DERs from all participants as well as the energy exchanges in the distribution network. A typical approach for energy trading is an auction-based mechanism where energy sellers submit their ask prices and buyers submit their bids then a system operator (SO) acts as an auctioneer that calculates a clearing price. The sellers who asked less than the clearing price and the buyers who bid more than the clearing price get their transactions fulfilled. Optimal bidding strategies were proposed in these studies \cite{liu2018novel, liu2019intraday, zhong2017auction, guerrero2018decentralized, purage2019cooperative, zhang2019two} to help trading participants in maximizing their own welfare or reducing costs in satisfying their demand. Iterative auction methods and multi-stage strategies were formulated to arrive at optimal bidding and selling prices. Frameworks using power matching algorithm were also introduced in \cite{qin2017flexible, qin2018automatic} with an added feature of checking the trading proposal feasibility to preserve network constraints. After receiving transaction offers, the SO evaluates them and curtails offers that violate network constraints and then releases a guide to aid participants in changing their offers. Auction-based trading may be simple and straightforward but issues may arise from this type of mechanism. The individualistic approach of auction-based mechanisms where each participant only maximizes its own welfare may not result to the overall optimal state of the system. Another issue for this is the price of electricity that may be driven by large entities with large assets during the auction process which might be unfair for participants with smaller assets therefore discouraging trading participation.

A better approach would be to apply coordination among participants during trading wherein information from participants are used to maximize the overall welfare of the system by solving for DER dispatch, power flows and transaction prices. With this, the optimal states of the system is achieved and entities with large assets can not drive the price of electricity for their benefit. This approach is commonly applied with a game theoretic-based method wherein trading participants are modeled as players and the game is modeled as a profit maximization problem. A Stackelberg game, where producers lead and consumers follow, is widely used in modeling energy trading problems \cite{zhang2018peer, liu2017energy, maharjan2013dependable, wang2014game}. This approach was also used in \cite{anoh2019energy} where they optimize both the cost for consumers and the utility for producers. Same method was used for modeling the problem in \cite{mediwaththe2017competitive} where a storage device coordinator first maximizes its revenue by trading with the grid. The prosumers with PV assets then play a non-cooperative repeated game following the strategy of the storage coordinator. The operator then coordinates with the prosumers to minimize the total community energy cost.  Another approach is by using Nash Bargaining theory where a unique solution is derived for all participants wherein one cannot increase its benefit without decreasing other participant's benefit \cite{gao2014game, kim2013bidirectional, li2018distributed, kim2019direct, wang2016incentivizing}. Using this approach, joint energy scheduling and trading were modeled in a single problem to incorporate energy exchange and optimal power flow (OPF) technique. The problem is decomposed into two phases where the OPF is solved first followed by the optimal prices for the energy trading transactions. 

Two issues, however, are observed from recent studies on game theoretic-based approaches in energy trading. First one is that most of these do not consider full AC network model in a distribution system due to the the non-convex and nonlinear nature of the AC OPF problem. Power flow models used are relaxed and linearized \cite{kim2019direct, wang2019incentivizing, wang2019incentive, paudel2018hierarchical, morstyn2018bilateral, lezama2018local} for easier computation but it may not be accurate enough to simulate energy trading in real world setting. From our previous works, the results using full AC model \cite{asarias2019distributed} of the network have been observed to be more realistic than a DC network model \cite{asarias2017resilient,asarias2018resilient} especially for distribution networks where voltage constraints must be considered in the power flow calculations. Second is the issue on fairness in terms of benefits and contribution among the trading participants. For some studies \cite{li2018distributed, wang2016incentivizing}, the energy cost savings derived by adopting the Nash Bargaining solution shall be divided equally among the participants regardless of the amount of energy traded . For example, a participant that contributed 1 kWh of energy in the trading process will get equal benefits or incentives with those who contributed 50 kWh or even larger amounts of energy. This clearly presents a fairness issue for the participants in the system. To address this, a sharing contribution rate concept \cite{kim2019direct, wang2019incentive, wang2019incentivizing} was introduced to promote fairness in the system by making sure that all participants have equal profits per unit of energy traded. However, this fairness concept integrated in a system where full physical network constraints are considered is yet to be investigated. We take a particular interest in investigating how fairness is achieved among the set of producers and set of consumers in a real world setup. 

This paper proposes an energy trading framework for distribution systems that aims to fill the gaps on current energy trading studies where the full AC network and power flow models are not considered and the concept of fairness among trading participants in a full AC network model is yet to be studied and investigated. The proposed framework is composed of three control stages. In the first control stage, a cost minimization problem is solved to compute the optimal cost for each agent when no trading happens. The optimal cost calculated in the first stage is fed to the second stage and will serve as the reference in solving a profit maximization problem to determine the optimal dispatch of DERs, voltage settings, initial transaction prices and initial payments/incentives of the participants during trading. These will then be the inputs for the last control stage where a benefit allocation algorithm is employed to solve for the optimal transaction price for each agent and re-adjust the payments/incentives based on each agent's contribution during trading to uphold fairness in the system. Full AC power flow models and network constraints are used in the optimization problems in each control stage. The key contributions of this paper are the following:

\begin{itemize}
  \item A transactive energy framework that incorporates non-convex and nonlinear AC power flow model while considering  full network and economic constraints and  ensuring fairness in the system. The AC power flow model considered in this study is not relaxed and not linearized, in contrast with recent studies mentioned earlier, to properly simulate the energy trading in a real world setting. With the proposed framework, fairness is ensured for all participants by ensuring equal profits for the set of consumers and producers with respect to the amount of energy they imported and exported considering power losses incurred during trading.
  \item Formulation of benefit allocation algorithm for solving the optimal transaction price and recalculation of payments and incentives of trading participants. Using the initial results of power flow and transaction prices, the incentives and payments are recalculated  to achieve fairness based on the trading contribution of each participant. 
\end{itemize}

The rest of this paper is structured as follows. The modeling of the transactive energy system in a distribution network is presented in Section II. In Section III, the proposed transactive energy framework is presented together with the formulation of the benefit allocation algorithm. Simulation results and numerical analysis are presented in Section IV and conclusion and future works are discussed in Section V.

\section{Transactive Energy System Modeling}
The overall model of transactive energy system deployed in a distribution network is presented in this section. In the first subsection, the distribution network system model that includes the power flow model used in this study is presented while models of microgrid and DERs together with the cost functions used in this study are introduced in the second subsection.

\subsection{Distribution Network System Model}
\begin{figure}[t]
    \centering
    \includegraphics[width=0.5\textwidth]{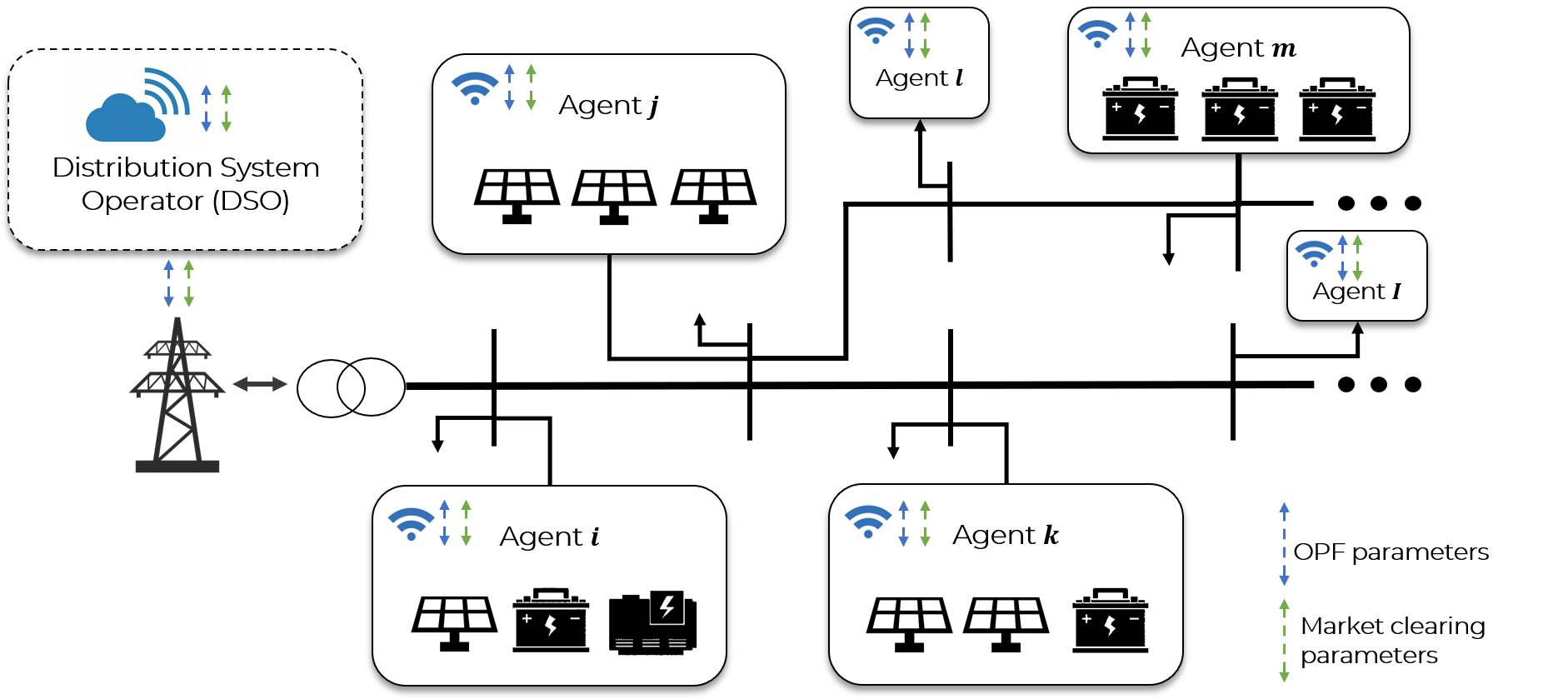}
    \caption{Transactive energy system diagram in a distribution network}
    \label{fig: systemdiagram}
\end{figure}

The distribution system is modeled as a radial network that can be represented by a graph $G(N,E)$ where $N$ is the number of buses in the network and $E \subseteq N \times N$  is the set of branches in the network. Each bus $i$ is considered as an agent with communication and computational capability that is either a  consumer, producer or prosumer (both consumer and producer). A set of microgrids $M \subseteq N$ with own mix of DERs is also considered in the system. A utility company connected to the radial network is modeled as a slack bus $1 \in N$ of $G$ as external power source of the system. The operational time horizon is divided into equal duration $\Delta t$ time slots (e.g. one hour, thirty minutes, five minutes, etc.). The overall system diagram of the distribution network is shown in Fig. \ref{fig: systemdiagram}. All agents exchange power flow and market clearing parameters with the distribution system operator (DSO) to compute for the optimal operational settings including the trading transactions in the network.

We consider a balanced three-phase system and for this part, we present a single-phase analysis of the network. The complex voltage of each bus $i \in N$ at time $t$ is denoted by $V_i(t) = e_i(t) + jf_i(t)$ where $e$ and $f$ are the real and imaginary parts of the voltage, respectively, and $j^2 = -1$. The line impedance for line $(i,j) \in E$ is denoted by $z_{ij} = r_{ij} + jx_{ij}$ where $r_{ij}$ and $x_{ij}$ are the line resistance and reactance, respectively. On the other hand, the admittance of line $(i,j) \in E$ is denoted by $y_{ij} = 1/z_{ij}$ and can also be expressed as $y_{ij} = g_{ij} + jb_{ij}$ where $g_{ij}$ and $b_{ij}$ are the conductance and susceptance, respectively. The complex current from bus $i$ to $j$ at time $t$ is denoted by $I_{ij}(t) = y_{ij}(V_i - V_j)$. We denote the complex conjugate of current $I_{ij}(t)$ as $I^{*}_{ij}(t)$ and define the complex power flow in line $(i,j) \in E$ at time $t$ as $S_{i,j}(t) = V_{i}(t)I^{*}_{ij}(t)$ and can also be expressed as $S_{ij}(t) = P_{ij}(t) + jQ_{ij}(t)$ where $P_{ij}(t)$ and $Q_{ij}(t)$ are the real and reactive power flows on line $(i,j)$ at time $t$, respectively.

The distribution system model considers all the constraints in the network namely power balance, generation capacity limits and voltage limits as shown in eq. \ref{P_energy_balance}-\ref{voltage_limit} for each bus $i \in N$. There are no relaxations made in the constraints to achieve full AC power flow analysis of a transactive energy system. This is also to ensure that no network constraints are violated during energy exchange among agents in the network and to achieve realistic results. For conciseness, the variable $t$ was removed from the equations. The generation $P_i^G$ and demand $P_i^D$ variables in eq. \ref{P_energy_balance} and \ref{Q_energy_balance} denote the total generation dispatch from all sources and total demand from all loads in each bus.

\begin{equation}
    \label{P_energy_balance}
    P_i^G-P_i^D=\sum_{j \in N} [e_i(g_{ij}e_j - b_{ij}f_j) + f_i(g_{ij}f_j + b_{ij}e_j)], \forall i 
\end{equation}

\begin{equation}
    \label{Q_energy_balance}
    Q_i^G-Q_i^D=\sum_{j \in N} [f_i(g_{ij}e_j - b_{ij}f_j) - e_i(g_{ij}f_j + b_{ij}e_j)], \forall i
\end{equation}

\begin{equation}
    \label{P_generation}
    \underline{P_i^G} \leq P_i^G \leq\overline{P_i^G}, \forall i 
\end{equation}

\begin{equation}
    \label{Q_generation}
    \underline{Q_i^G} \leq Q_i^G \leq\overline{Q_i^G}, \forall i
\end{equation}

\begin{equation}
    \label{voltage_limit}
    \underline{V_i} \leq e_i^2+f_i^2 \leq \overline{V_i},\forall i
\end{equation}

\subsection{DER Models and Cost Functions}
Different types of entities connected to each bus can be a producer, consumer or prosumer of energy. Pure load buses represent consumer entities while buses with pure generation represents renewable energy-based power plants such as PV and wind farms or fuel-based power plants that are connected to the distribution system. In this study, microgrids are assumed to have their own set of power generation (renewable and/or fuel-based) that services a load demand. The overall objective of the system before trading is to minimize the generation costs in satisfying the system load demand. With this, optimal dispatch of generation sources as well as power purchased from the grid must be determined while satisfying network constraints. The following shows the cost functions of energy resources considered in the system.

\subsubsection{Imported Power}
Each bus can purchase/sell deficit/excess power from/to the grid. We denote the imported power of bus $i$ at time $t$ to be $P_{im,i}(t)$. A negative $P_{im,i}(t)$ means bus $i$ exported power at time $t$. The purchasing price from utility is denoted by $u_b(t)$ (\textdollar/kWh) while the selling price to the utility is denoted by $u_s(t)$ (\textdollar/kWh). The cost function for importing and exporting power is shown in (\ref{cost_importedpower}) where $\pi_i(t)$ is the transaction price at time $t$. For the case of no trading, $\pi_i(t) = u_b(t)$ when bus $i$ imports power from the grid and $\pi_i(t) = u_s(t)$ when bus $i$ exports power to the grid. On the other hand, during trading, $\pi_i(t)$ is determined by solving the profit maximization problem that is to be discussed in the next section.

\begin{equation}
    C_{im,i}(t) = P_{im,i}(t)\pi_i(t)\Delta t
    \label{cost_importedpower}
\end{equation}

\subsubsection{Battery}
Let $P_{b,i}(t)$ be the power discharged from the battery of bus $i$ at time $t$. A negative $P_{b,i}(t)$ means that the battery charged at time $t$. The energy in the battery at time $t$ is denoted by $E_{b,i}(t)$ and follows (\ref{battery_energy}) where $\eta_{d,i}$ and $\eta_{c,i}$ are the discharging and charging efficiency of the battery, respectively, with values from 0 to 1.

\begin{equation}
    E_{b,i}(t+1) = \begin{cases} 
          E_{b,i}(t) - \eta_{d,i} P_{b,i}(t)\Delta t & P_{b,i}(t) \geq 0 \\
          E_{b,i}(t) + \eta_{c,i} P_{b,i}(t)\Delta t & P_{b,i}(t) < 0
       \end{cases}
    \label{battery_energy}
\end{equation}

A battery's state of charge (SoC) must also be constrained to avoid too much stress and prolong battery lifespan as shown in (\ref{battery_SOC}) where ${SoC}_i^{min}$ and ${SoC}_i^{max}$ denote the minimum and maximum SoC and $\overline{E}_{b,i}$ is the maximum energy capacity of battery $i$.

\begin{equation}
    {SoC}_i^{min} \leq \frac{E_{b,i}(t)}{\overline{E}_{b,i}} \leq {SoC}_i^{max}
    \label{battery_SOC}
\end{equation}

The power dispatch from the battery is constrained and limited to maximum charging and discharging capacities denoted by $P_{b,i_c}^{max}$ and $P_{b,i_d}^{max}$, respectively, and is given by (\ref{battery_chargelimit})-(\ref{battery_dischargelimit}).

\begin{equation}
    0 \leq |P_{b_i}(t)| \leq P_{b,i_c}^{max} \quad P_{b,i}(t) \geq 0 
    \label{battery_chargelimit}
\end{equation}

\begin{equation}
    0 \leq |P_{b_i}(t)| \leq P_{b,i_d}^{max} \quad P_{b,i}(t) < 0 
    \label{battery_dischargelimit}
\end{equation}

The degradation cost coefficient of the battery is denoted by $c_{d,i}$ (\textdollar/kWh) therefore the cost of operating a battery at time $t$ is given by (\ref{cost_battery}). 

\begin{equation}
    C_{b,i}(t) = c_{d,i}|P_{b,i}(t)|\Delta t
    \label{cost_battery}
\end{equation}

\subsubsection{Fuel-Based Generator}
The dispatched power from a fuel-based generator at time $t$, in this case a diesel generator (DG), is denoted by $P_{dg,i}(t)$ and is constrained by its minimum and maximum power dispatch capacity as given by (\ref{dg_limits}).

\begin{equation}
    {P}_{dg,i}^{min} \leq P_{dg,i} \leq {P}_{dg,i}^{max}
    \label{dg_limits}
\end{equation}

We model the cost function of a DG using a quadratic cost function as shown in (\ref{cost_dg}) where $a_{dg,i}  (\$/kWh^2), b_{dg,i} (\$/kWh), c_{dg,i} (\$)$ are the DG cost coefficients.

\begin{equation}
    C_{dg,i}(t) = a_{dg,i} P_{dg,i}(t)^2 + b_{dg,i} P_{dg,i}(t) + c_{dg,i}
    \label{cost_dg}
\end{equation}

\subsubsection{Total Cost Function}
Each bus may also have its own renewable energy (RE) resources and we define the available renewable power at time $t$ as $P_{re,i}(t)$. We only consider in this study the generation cost of RE sources which is zero. Using the models for energy sources, we define the active power balance in each bus using (\ref{powerbalance_activepower}) where $P_i^D(t)$ is the load demand at time $t$. The right hand side of the equation is the sum of all power from different resources at time $t$ and is denoted as $P_i^G(t)$ which must be equal to the load demand shown at the left hand side. All loads are assumed to be fixed for this study.

\begin{equation}
    P_i^D(t) = P_{im,i}(t) + P_{b,i}(t) + P_{dg,i}(t) + P_{re,i}(t)
    \label{powerbalance_activepower}
\end{equation}

We also consider the cost of power losses incurred by each bus when importing power from the grid. Let $P_{l,i}(t)$ be the power loss incurred by bus $i$ at time $t$ which can be calculated by (\ref{powerloss}) where $i \sim j$ denotes buses connected to bus $i$. 

\begin{equation}
    P_{l,i}(t) = \sum_{i \sim j } S_{ij}(t) + S_{ji}(t)
    \label{powerloss}
\end{equation}

The cost of power loss is the amount of power loss multiplied by the transaction price at time $t$ as shown in (\ref{cost_powerloss}). During trading, the price will be determined by solving the profit maximization problem, while on the case of no trading, $\pi_i(t) = u_b(t)$.

\begin{equation}
    C_{l,i}(t) = P_{l,i}(t)\pi_i(t)\Delta t
    \label{cost_powerloss}
\end{equation}

The cost function of each bus denoted by ${C_{i}}(t)$ is the sum of generation cost of the DERs and costs of imported power and power loss as shown in (\ref{costfunction_no_trading}). 

\begin{equation}
    {C_{i}}(t) = C_{im,i}(t) + C_{b,i}(t) + C_{dg,i}(t) + C_{l,i}(t)
    \label{costfunction_no_trading}
\end{equation}

Note that the total cost function for each bus has the same form for both cases (no trading, with trading) but differs in the cost of imported or exported power due to the difference in pricing. With this, we denote the total cost for bus $i$ at time $t$ as $\bar{C}_i(t)$ for the case of no trading and $\tilde{C}_i(t)$ for the case of trading in the system.

\section{Three-Stage Transactive Energy Framework}
\begin{figure}[]
    \centering
    \includegraphics[width=0.48\textwidth]{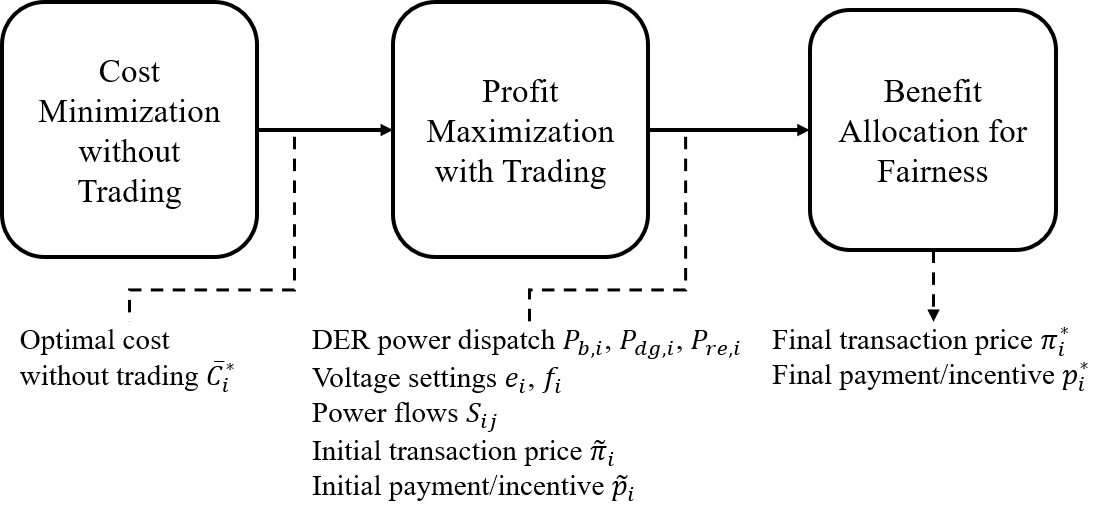}
    \caption{The three control stages of proposed energy trading framework}
    \label{fig: framework}
\end{figure}

The proposed transactive energy trading framework is discussed in this section. There are three control stages in the framework as shown in Fig. \ref{fig: framework} wherein fairness among agents participating in trading is ensured while satisfying the network constraints in the system. The following subsections will discuss each control stage in the proposed trading framework.

\subsection{Cost Minimization without Trading}
The first stage of the framework deals with a typical cost minimization problem wherein the overall objective of the system is to minimize the power generation costs as shown in (\ref{objfunction_no_trading}) while satisfying demand and network constraints (\ref{P_energy_balance}) - (\ref{voltage_limit}) where the variables to be solved are $\mathbf{\bar{P}_i} = (\bar{P}_{im,i}(t), \bar{P}_{b,i}(t), \bar{P}_{dg,i}(t), \bar{P}_{re,i}(t))$, $\mathbf{\bar{Q}_i} = (\bar{Q}_{im,i}(t), \bar{Q}_{b,i}(t), \bar{Q}_{dg,i}(t), \bar{Q}_{re,i}(t))$, $\mathbf{\bar{e}_i} = (\bar{e}_i(t))$ and $\mathbf{\bar{f}_i} = (\bar{f}_i(t))$ for all $i \in N$.

\begin{equation}
    \underset{\mathbf{\bar{P}_i}, \mathbf{\bar{Q}_i}, \mathbf{\bar{e}_i}, \mathbf{\bar{f}_i}}{\text{min}} \sum_{i \in N}  \bar{C_{i}}(t)
    \label{objfunction_no_trading}
\end{equation}

\hspace{0.5cm}  $s.t. \hspace{0.6cm} (\ref{P_energy_balance}) - (\ref{voltage_limit})$
\vspace{0.2cm}

Solving (\ref{objfunction_no_trading}) will give the optimal dispatch of DERs, amount of power to be imported or exported, values of voltage at each bus and the power flows between buses in the system at time $t$ without trading transactions among agents. We denote the optimal cost of each bus without trading as $\bar{C}_{i}^{*}(t)$ that will be used in the next stage of the trading framework.

\subsection{Profit Maximization with Trading}
Upon solving for the least cost settings of the system in the previous stage, we introduce another  control stage that aims to maximize the benefits of the whole system. Essentially, this control stage aims to find the transaction prices for buses that imported or exported power which will reduce the total cost of power to meet the demand and satisfy the network constraints in the system. We denote the profit of bus $i$ at time $t$ as $\varrho_i(t)$ that is the difference between the cost without trading and the cost with trading as shown in (\ref{profit}). With this, profit must always be positive i.e. the cost incurred by bus $i$ when trading energy with other agents  must always be cheaper compared when there is no trading with other agents.

\begin{equation}
    \varrho_i(t) = \bar{C}_i^{*}(t) - \Tilde{C}_i(t)
    \label{profit}
\end{equation}

Transaction pricing bounds must be taken into careful consideration to be able to encourage participants in trading. With this, buying price of energy from other participants must be lower than the buying price from utility. On the other hand, the selling price for those who exported power must be higher than the selling price to utility. These pricing bounds shown in (\ref{pricing_bounds}) ensure that those participating in the trading process automatically benefits from the energy exchange with other participants as compared to transacting with the utility.

\begin{equation}
    u_s(t) \leq \pi_i(t) \leq u_b(t)
    \label{pricing_bounds}
\end{equation}

Another trading constraint that must be considered is that the payments and incentives of agents participating in trading must be equal to zero since the payment of one agent for the energy imported is the incentive of the agent that exported. Cost of power loss is included to account for line losses during trading. The additional trading constraint is shown in (\ref{sum_PimpPloss}) where payment at time $t$ is denoted as $\delta_{i}(t)$. Negative $\delta_{i}(t)$ values represent incentives for exported power. The cost of power loss can also be considered as the grid-access fee in the system or the cost in using the utility's infrastructure during trading.

\begin{equation}
    \sum_{i \in N} \delta_{i}(t) =  \left( \Tilde{P}_{im,i}(t) + \Tilde{P}_{l,i}(t) \right) \pi_i(t)   = 0\
    \label{sum_PimpPloss}
\end{equation}

Given the equation for benefit or profit for each bus, we formulate the optimization problem for this control stage. We consider the welfare or benefit of each bus to be represented by the profit incurred during trading. The problem can then be formulated as a profit maximization problem as shown in (\ref{obj_function_trading}) where the variables to be solved are $\mathbf{\Tilde{P}_i} = (\Tilde{P}_{im,i}(t), \Tilde{P}_{b,i}(t), \Tilde{P}_{dg,i}(t), \Tilde{P}_{re,i}(t))$, $\mathbf{\Tilde{Q}_i} = (\Tilde{Q}_{im,i}(t), \Tilde{Q}_{b,i}(t), \Tilde{Q}_{dg,i}(t), \Tilde{Q}_{re,i}(t))$, $\mathbf{\Tilde{e}_i} = (\Tilde{e}_i(t))$, $\mathbf{\Tilde{f}_i} = (\Tilde{f}_i(t))$ and $\mathbf{\Tilde{\Pi}_i} = (\Tilde{\pi}_i(t))$  for all $i \in N$.

\begin{equation}
    \underset{\mathbf{\Tilde{P}_i}, \mathbf{\Tilde{Q}_i}, \mathbf{\Tilde{e}_i}, \mathbf{\Tilde{f}_i}, \mathbf{\Tilde{\Pi}_i}}{\text{max}} \sum_{i \in N}  \varrho_i(t)
    \label{obj_function_trading}
\end{equation}

\hspace{0.5cm}  $s.t. \hspace{0.6cm} (\ref{P_energy_balance}) - (\ref{voltage_limit}), (\ref{pricing_bounds}), (\ref{sum_PimpPloss})$
\vspace{0.2cm}

The initial transaction price for buses that traded power is determined upon solving the profit maximization problem as well as the payment and incentives for the transactions. However, these payments and incentives do not automatically imply fairness for all agents that participated in the trading. Therefore, another control layer is added to ensure fairness for all the trading participants in the system.

\subsection{Benefit Allocation for Fairness}

\begin{algorithm}[t]
    \caption{Benefit Allocation for DSO}
    \SetAlgoLined
    \SetKwInOut{Input}{Input}\SetKwInOut{Output}{Output}
    
    \Input{Imported/Exported power $P_{im,i}(t)$, initial transaction payments/incentives $\delta_{i}(t)$, initial transaction price $\pi_{i}(t)$, $\forall{i}$}
    \Output{Optimal transaction price $\pi_i^{*}(t)$, optimal transaction payments and incentives $\delta_{i}^{*}(t)$, $\forall{i}$}
    
    \For{$t=1$ \KwTo $T$}{
    Get the total exported and imported power at time $t$.\\
    Compute trading contribution rate $r_{t,i}$ of each bus $i$ with respect to total imported and exported power.\\
    Get sum of incentives for exported power and payments for imported power.\\
    Recalculate incentives and payment $\delta_{i}^{*}(t)$ of each bus $i$ with respect to trading contribution rate $r_{t,i}$.\\
    Calculate optimal transaction price $\pi_i^{*}(t)$ using recalculated incentives and payment with respect to the exported and imported power.\\
    }
    \label{algo: ben_alloc}
\end{algorithm}

The last control stage of the proposed framework aims to address the issue on fairness in energy trading. The benefit allocation algorithm shown in Algorithm \ref{algo: ben_alloc} takes into account the actual payment and incentives initially calculated in the previous stage. Previous studies \cite{kim2019direct, wang2019incentive} have modeled fairness without including the power losses which must be considered in deriving the profit per unit energy transacted to achieve more realistic trading results. In addition, the proposed algorithm ensures that profits per unit energy for the set of producers/consumers to be equal with respect to the amount of exported/imported energy they contributed during trading. We introduce trading contribution rate $r_{t,i}$ that denotes the share or contribution of each bus to the overall imported or exported power in the system as shown in (\ref{trading_contribution_rate}). The incentives/payments for the transactions made will then be recalculated according to $r_{t,i}$ to calculate appropriate profits that ensure fairness to all the trading participants. The optimal transaction price $\pi_i^{*}(t)$ and the optimal transaction payments/incentives $\delta_{i}^{*}(t)$ for each bus can then be calculated using their respective $r_{t,i}$.

\begin{equation}
    r_{t,i} = \frac{|P_{im,i}|}{\sum_{j \in N}|P_{im,j}|}
    \label{trading_contribution_rate}
\end{equation}

\begin{table}[]
    \centering
    \caption{Simulation Parameters}
    \begin{tabular}{@{}cclll@{}}
    \toprule
    \textbf{Parameter}             & \multicolumn{4}{c}{\textbf{Value}}   \\ \midrule
    Time steps per day             & \multicolumn{4}{c}{48}               \\
    Battery capacity               & \multicolumn{4}{c}{1000 $kWh$}         \\
    Battery charging efficiency    & \multicolumn{4}{c}{0.9}              \\
    Battery discharging efficiency & \multicolumn{4}{c}{0.9}              \\
    Battery degradation cost       & \multicolumn{4}{c}{0.1 $\$/kWh$}        \\
    Maximum Battery Power          & \multicolumn{4}{c}{500 $kW$}           \\
    SOC range                      & \multicolumn{4}{c}{{[}0.4, 0.9{]}}   \\
    Maximum DG Power               & \multicolumn{4}{c}{1000 $kW$}          \\
    Voltage range (p.u.)           & \multicolumn{4}{c}{{[}0.95, 1.05{]}} \\
    $a_{dg,i}$                           & \multicolumn{4}{c}{2.45e-5 $\$/kWh^2$}          \\
    $b_{dg,i}$                              & \multicolumn{4}{c}{0.1833 $\$/kWh$}           \\
    $c_{dg_i}$                              & \multicolumn{4}{c}{26.235 $\$$}           \\ \bottomrule
    \end{tabular}
    \label{tab: sim_param}
\end{table}

\begin{figure}[] 
    \centering
    \includegraphics[width=0.48\textwidth]{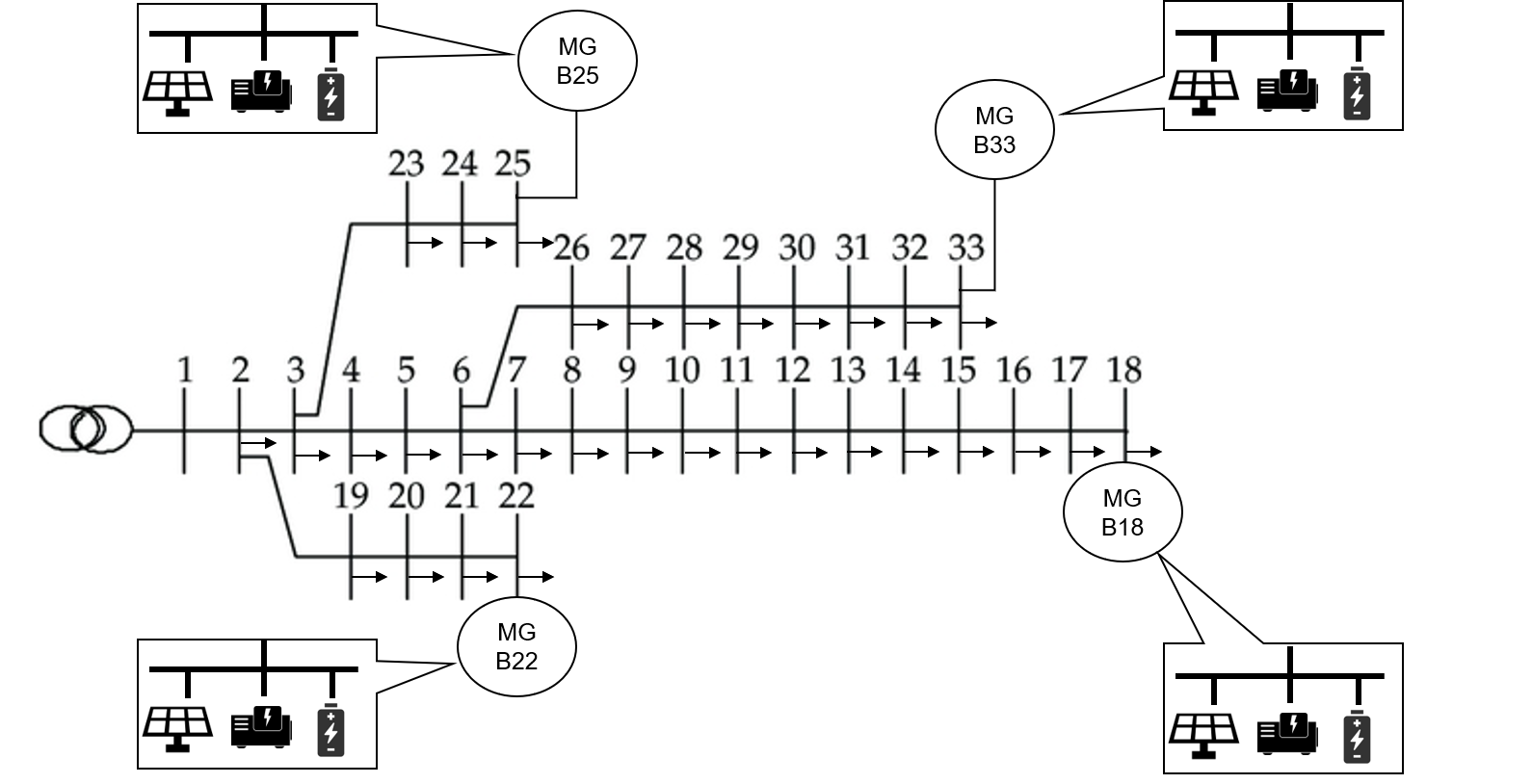}
    \caption{IEEE 33-Bus network used in the simulation}
    \label{fig:simulation_network}
\end{figure}

\section{Simulation and Analysis}
This section presents the simulation results of trading in a distribution system using the proposed trading framework. The first subsection presents the details of the simulation and network setup while numerical analysis of trading results are discussed in the second subsection.

\subsection{Simulation Setup}
Shown in Fig. \ref{fig:simulation_network} is the distribution network used in the simulation with multiple microgrids integrated having their own respective mix of DERs. All buses except the slack bus have their own load demand. The purchasing price from utility is based from the hourly time-of-use (ToU) pricing from California Independent System Operator (CAISO) \cite{shi2014distributed}. The half-hour purchasing price used in the simulation is derived by taking the average of two consecutive hourly prices. The selling price of excess energy to the utility is assumed to be half of the purchasing price. The rest of the parameters used in the simulation are in Table \ref{tab: sim_param}. The modeling and simulations were performed in MATLAB. The computer used for the experiments has a CPU Intel Core i7-8750H 2.20 GHz and 32 GB of RAM.


\subsection{Results of Trading}
\begin{figure}[t] 
    \centering
    \includegraphics[width=0.5\textwidth]{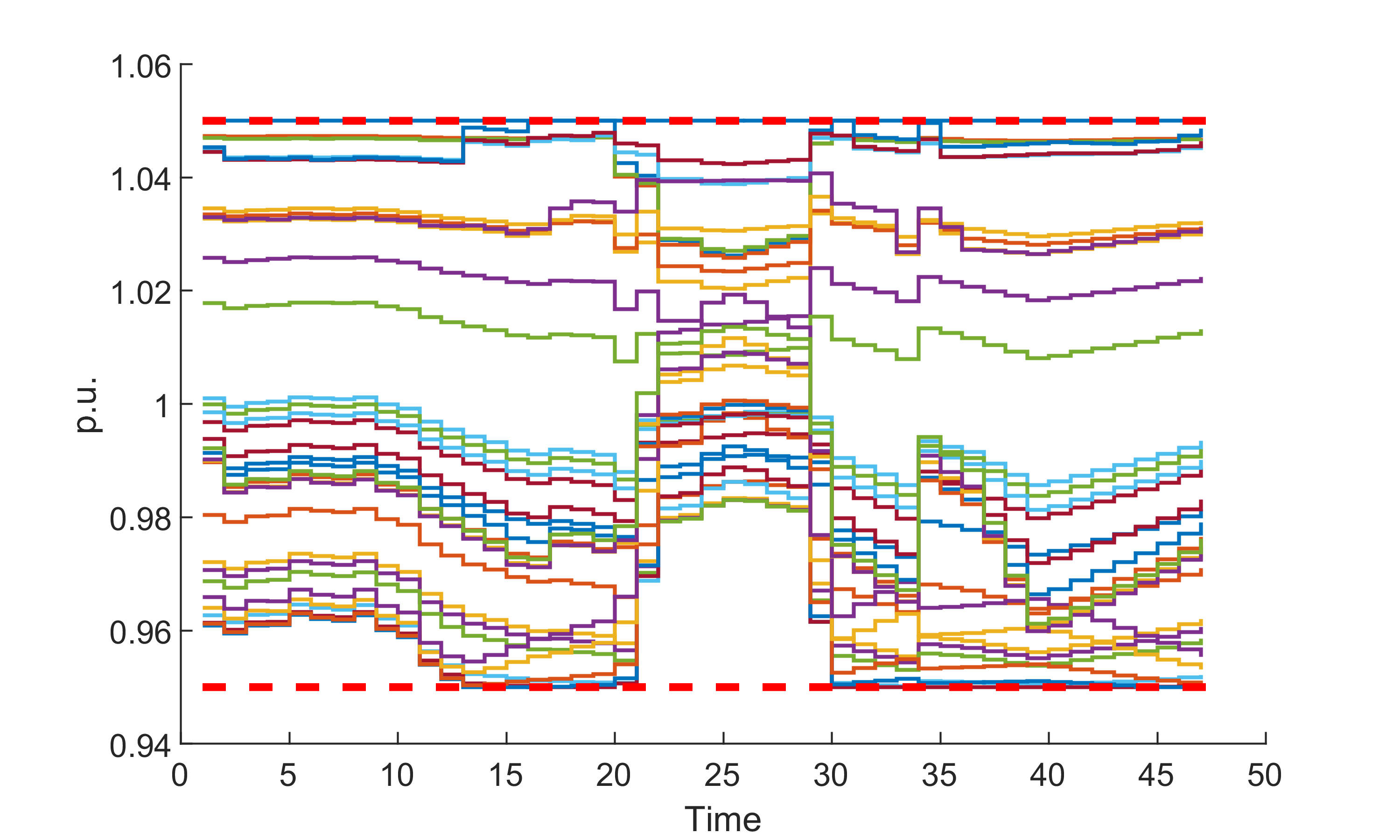}
    \caption{Voltage of buses during trading}
    \label{fig:voltage}
\end{figure}

The results presented in this section are the outputs of the second control stage of the proposed framework. Using the costs for no trading case as the reference, the profit of the whole system is maximized by finding the dispatch of DERs and transaction prices for imported and exported power by each bus or agent. Fig. \ref{fig:voltage} shows that voltage setting for each bus in the network is within the specified range and no voltage constraint was violated. This is an important consideration in distribution networks since deviations from allowable range could bring issues for end-use customers.

\begin{figure}[] 
    \centering
    \includegraphics[width=0.45\textwidth]{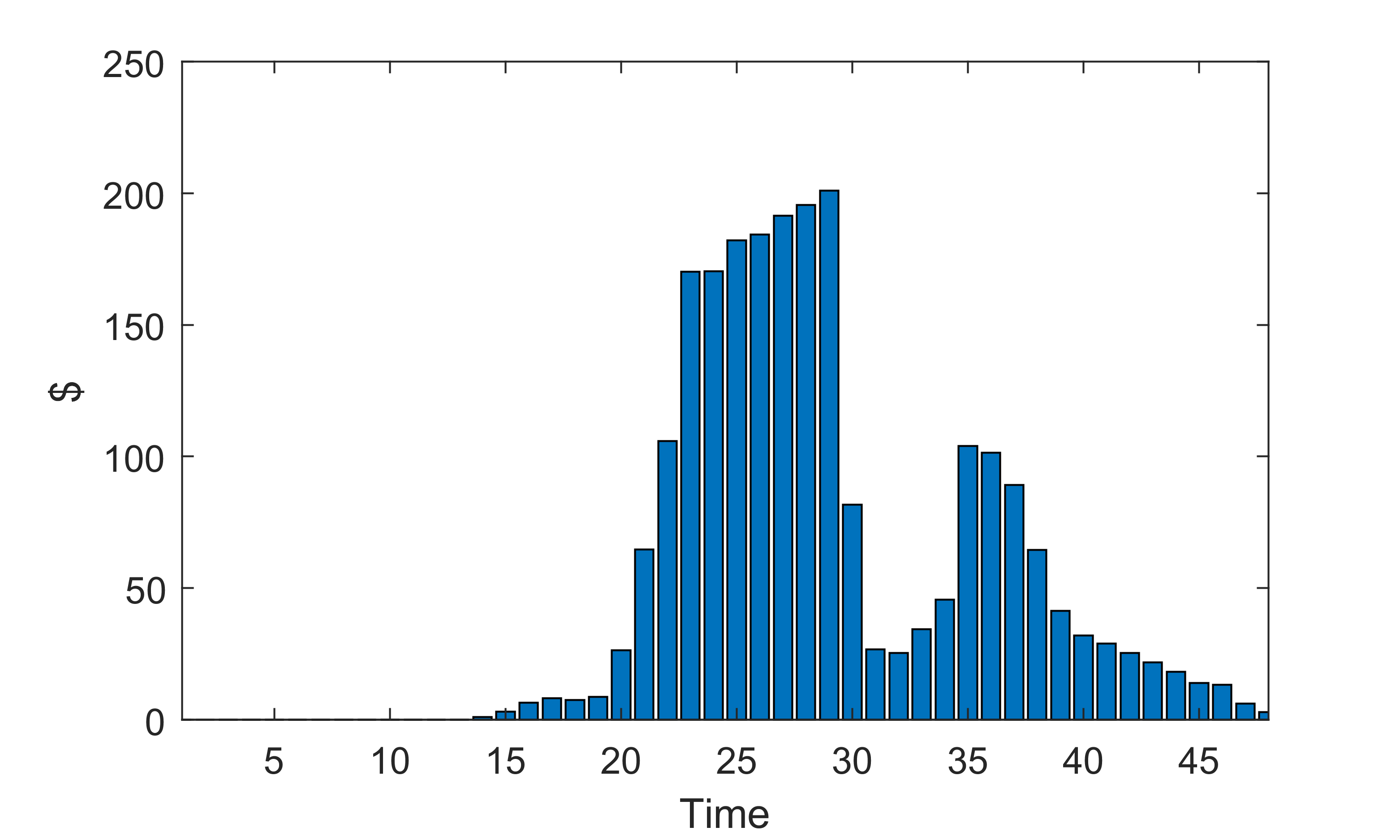}
    \caption{Total profit of the system}
    \label{fig:profit_total}
\end{figure}

\subsubsection{Cost Reduction}

\begin{figure}[t] 
    \centering
    \includegraphics[width=0.45\textwidth]{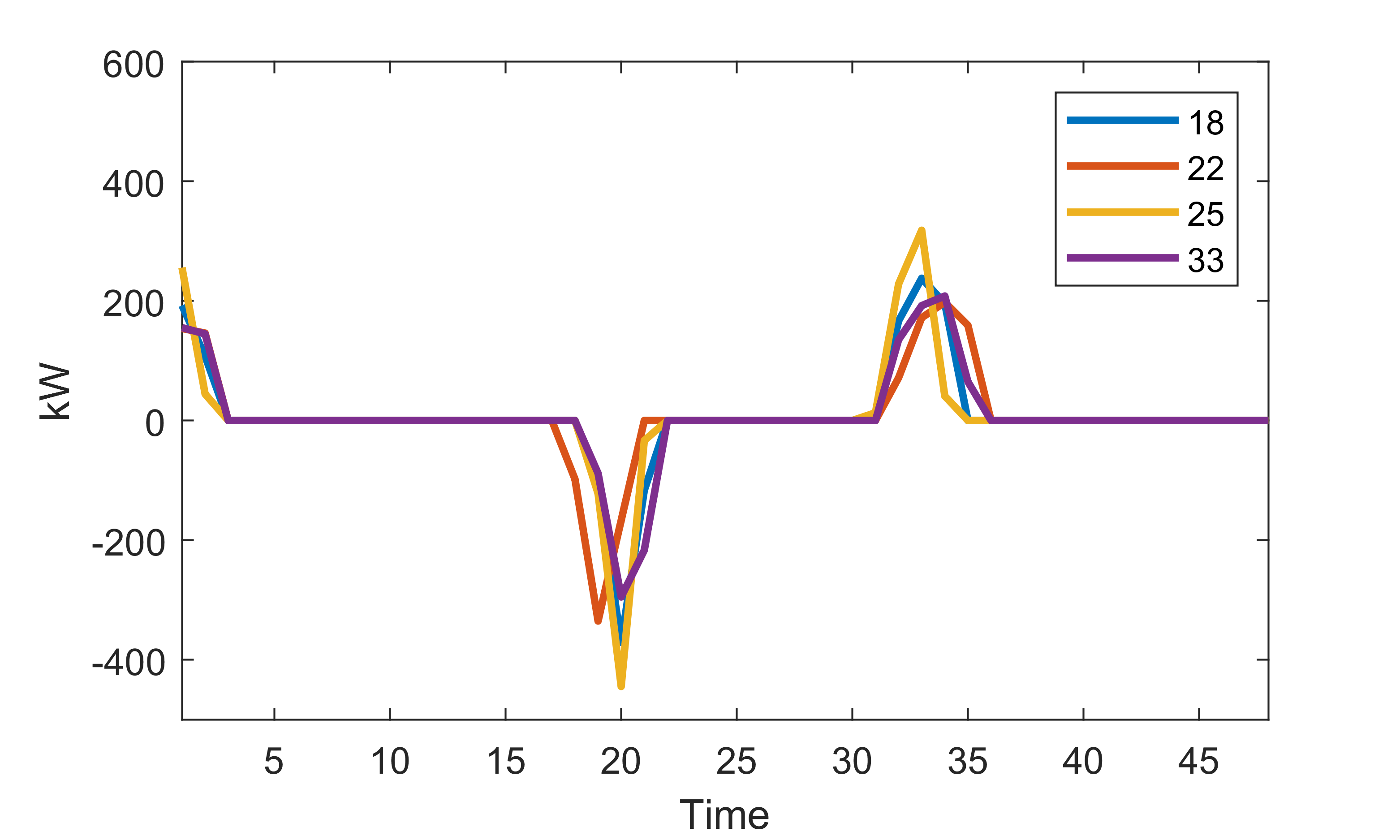}
    \caption{Battery power dispatch}
    \label{fig:batt_dispatch}
\end{figure}

\begin{figure}[t] 
    \centering
    \includegraphics[width=0.45\textwidth]{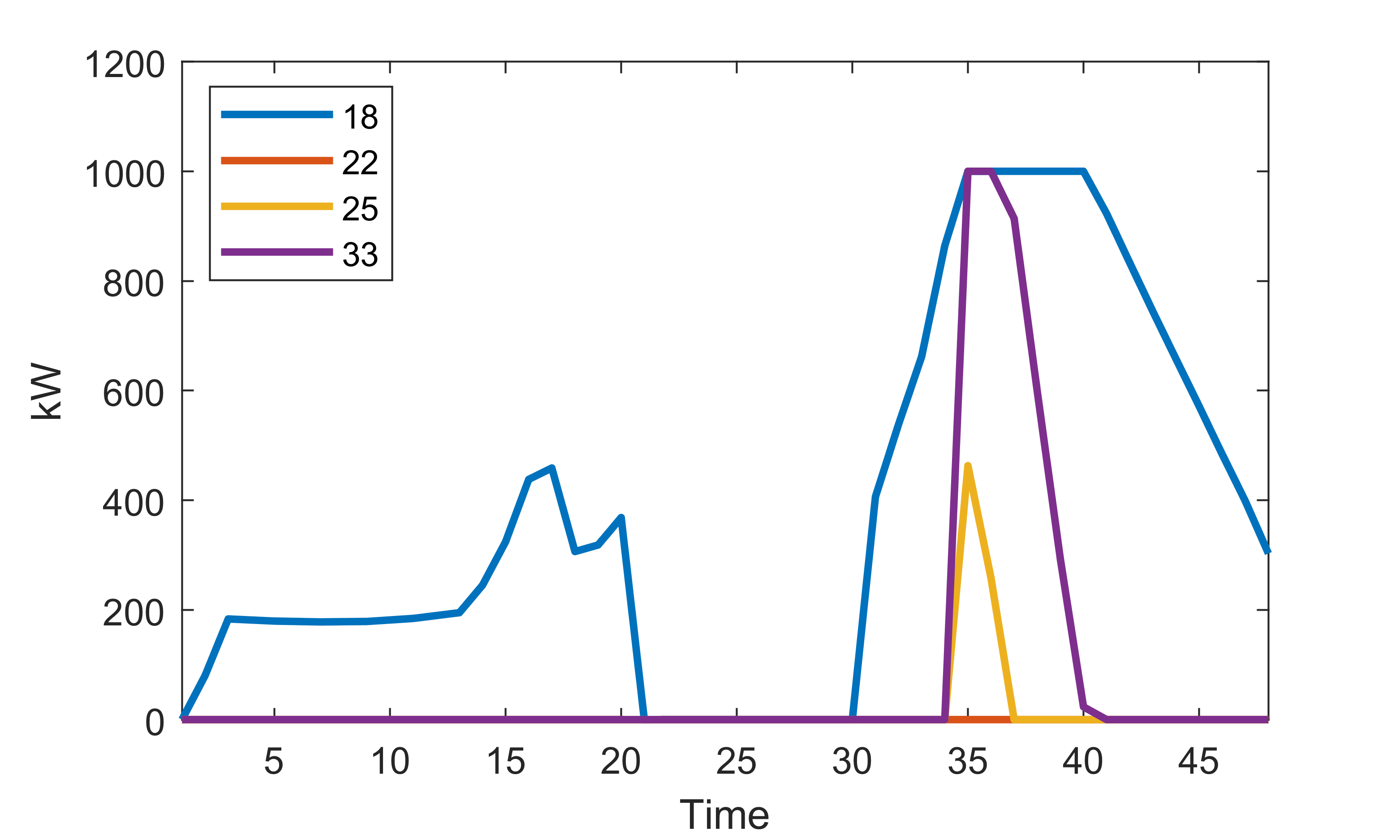}
    \caption{Diesel generator power dispatch}
    \label{fig:dg_dispatch}
\end{figure}

All agents participating in trading are being incentivized through a reduction in their power generation costs which is the main benefit of integrating trading in the operation of distribution network systems.  Shown in Fig. \ref{fig:profit_total} is the total profit for all agents in the system at each time step for a one day simulation. Notice that the majority of the incentives occur at the middle of the day wherein there is excess energy coming from PV resources. We also note that during this window, battery power is negative which indicates charging from cheaper excess energy from PV as shown in Fig. \ref{fig:batt_dispatch}. In addition, power coming from the utility also decreased due to self consumption of solar energy and export of excess by buses with PV resources. Another window for increase in profit can be observed at time $t=36$ or around night time wherein buying price from utility starts to increase due to increased demand. Around this window, we can observe dispatch from batteries and DG (Fig. \ref{fig:dg_dispatch}) from the microgrids to support the load demand and to provide cheaper source of energy as compared to power coming from the utility. The microgrids are exporting at this window to exploit the trading process by providing cheaper energy to other buses for more incentives. Essentially, all participants benefit from the trading process since consumers are buying cheaper energy while producers are selling excess energy for a higher rate as compared to the utility rate resulting to cost reduction and overall maximized welfare in the system. We note that the utility makes no profit during the trading process but a grid access fee is charged to every participant proportional to power loss incurred during trading as payment for the use of the utility's infrastructure.


\subsubsection{Balance in Cash Flow}
\begin{figure}[] 
    \centering
    \includegraphics[width=0.45\textwidth]{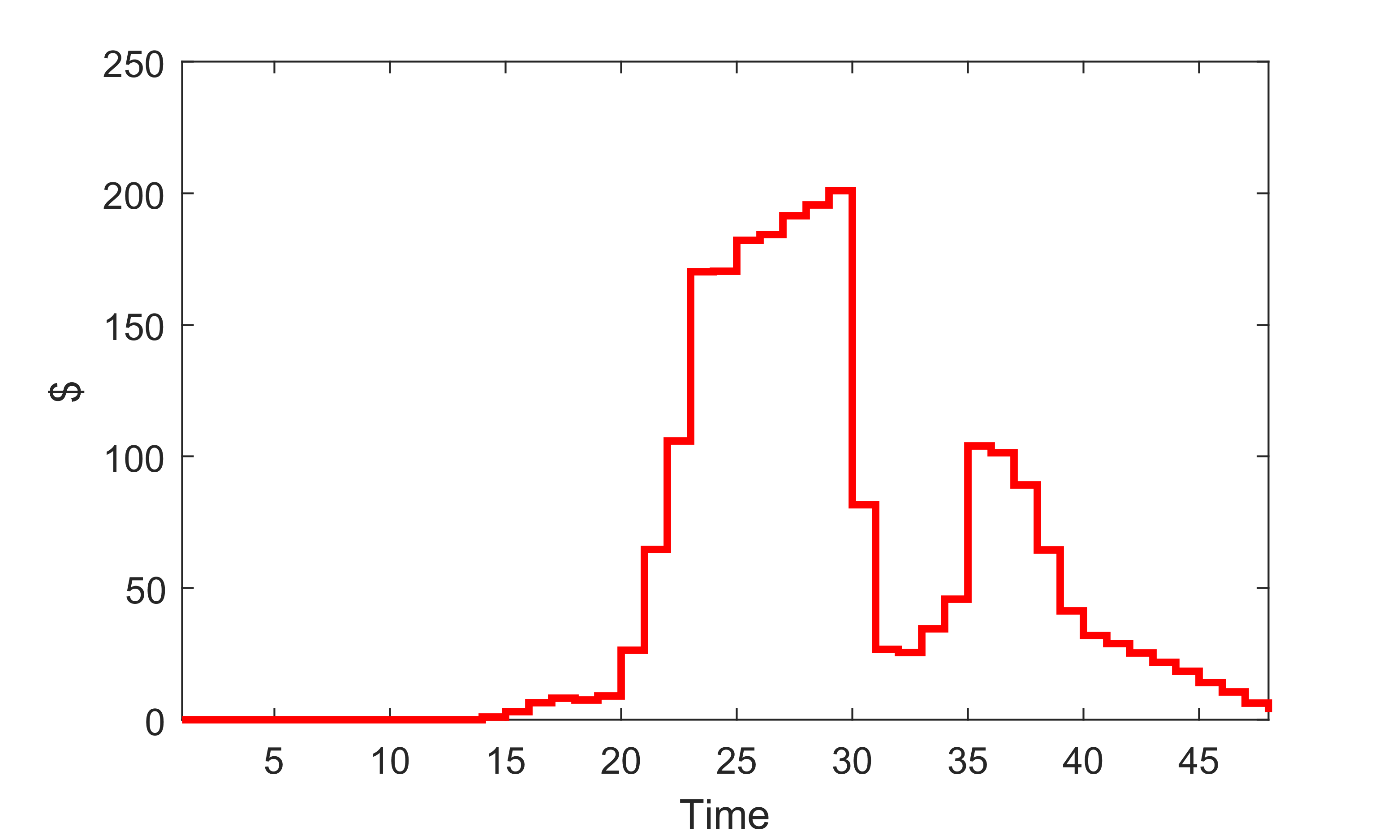}
    \caption{Imbalance in cash flow for no trading case}
    \label{fig:imbalance}
\end{figure}

\begin{figure}[] 
    \centering
    \includegraphics[width=0.45\textwidth]{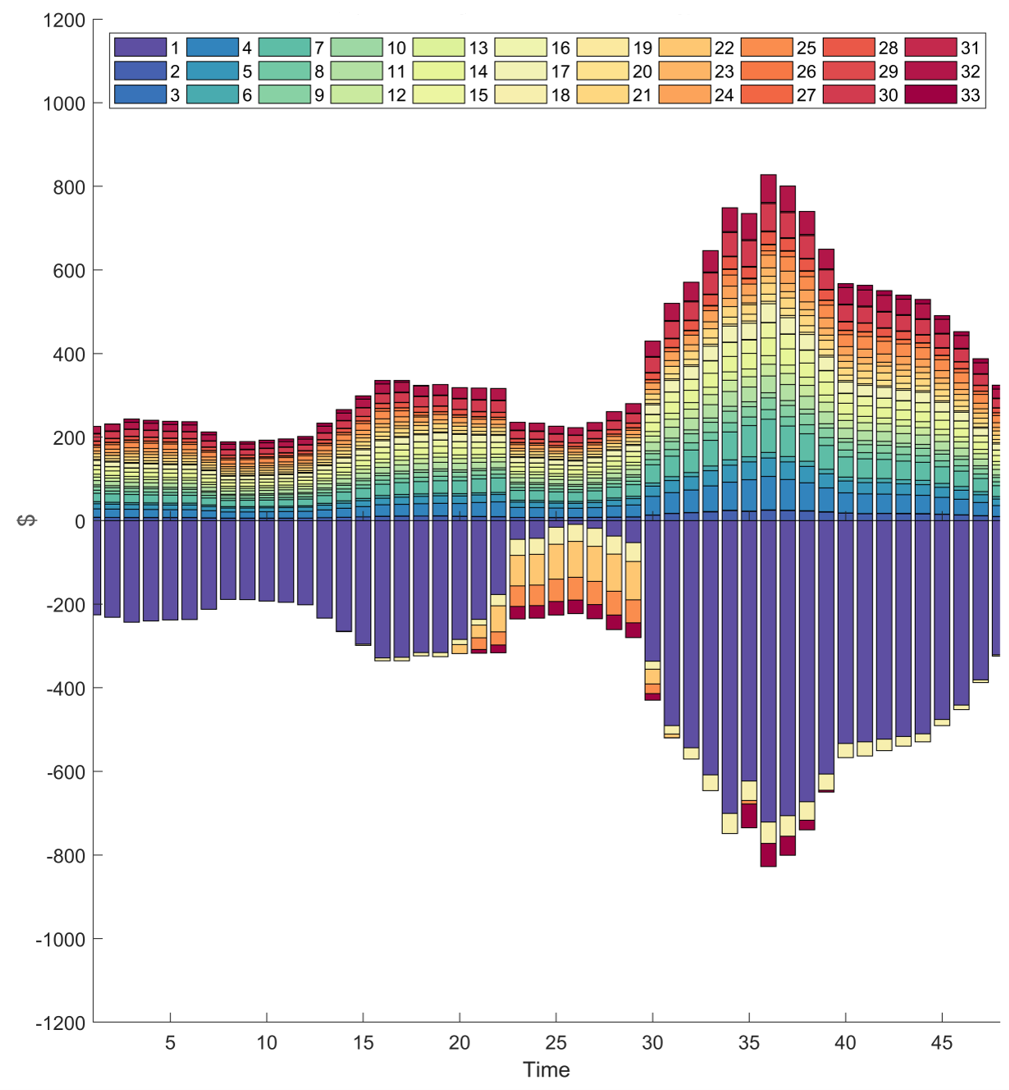}
    \caption{Cash flow in the system during trading}
    \label{fig:cash flow}
\end{figure}

Another benefit of trading is balance of cash flow for the system. This means that the sum of payments and incentives among participants must be zero or balance out after the trading process. For the traditional transaction of energy with the utility, there are instances that cash flow in the system does not balance out wherein payments for the procured energy are greater than the supposed incentive for the utility i.e. payments to utility can be greater than what it is supposed to receive. This is due to the fixed pricing scheme of the utility company as opposed to the the proposed trading framework wherein the optimal price is solved per transaction to maintain the balance in cash flow. Shown in Fig. \ref{fig:imbalance} are the imbalances in cash flow when no trading occurs. An interesting thing to observe is that these imbalances are the same with the profit of the system during trading as seen in Fig. \ref{fig:profit_total}. This shows that during trading, the imbalances in cash flow are resolved and distributed among agents to satisfy the constraint on zero sum of payments and incentives which is done by calculating the transaction price for each bus. This ensures there is no excess or deficit in  payments and incentives for each bus. A balanced cash flow for the system during trading is shown in Fig. \ref{fig:cash flow}. We can see from this graph that the payments (positive values) are equal with the incentives (negative values) at each time step.

\subsection{Benefit Allocation}
The outputs of the second control stage are the optimal dispatch of DERs including the imported/exported power, initial transaction price and initial transaction payments/incentives. These results, however, do not guarantee fairness for all participants in the trading process. This section presents the results of integrating the last control stage in the proposed framework for ensuring fairness for all participants in trading.

\subsubsection{Effect in Price}
\begin{figure}[] 
    \centering
    \includegraphics[width=0.45\textwidth]{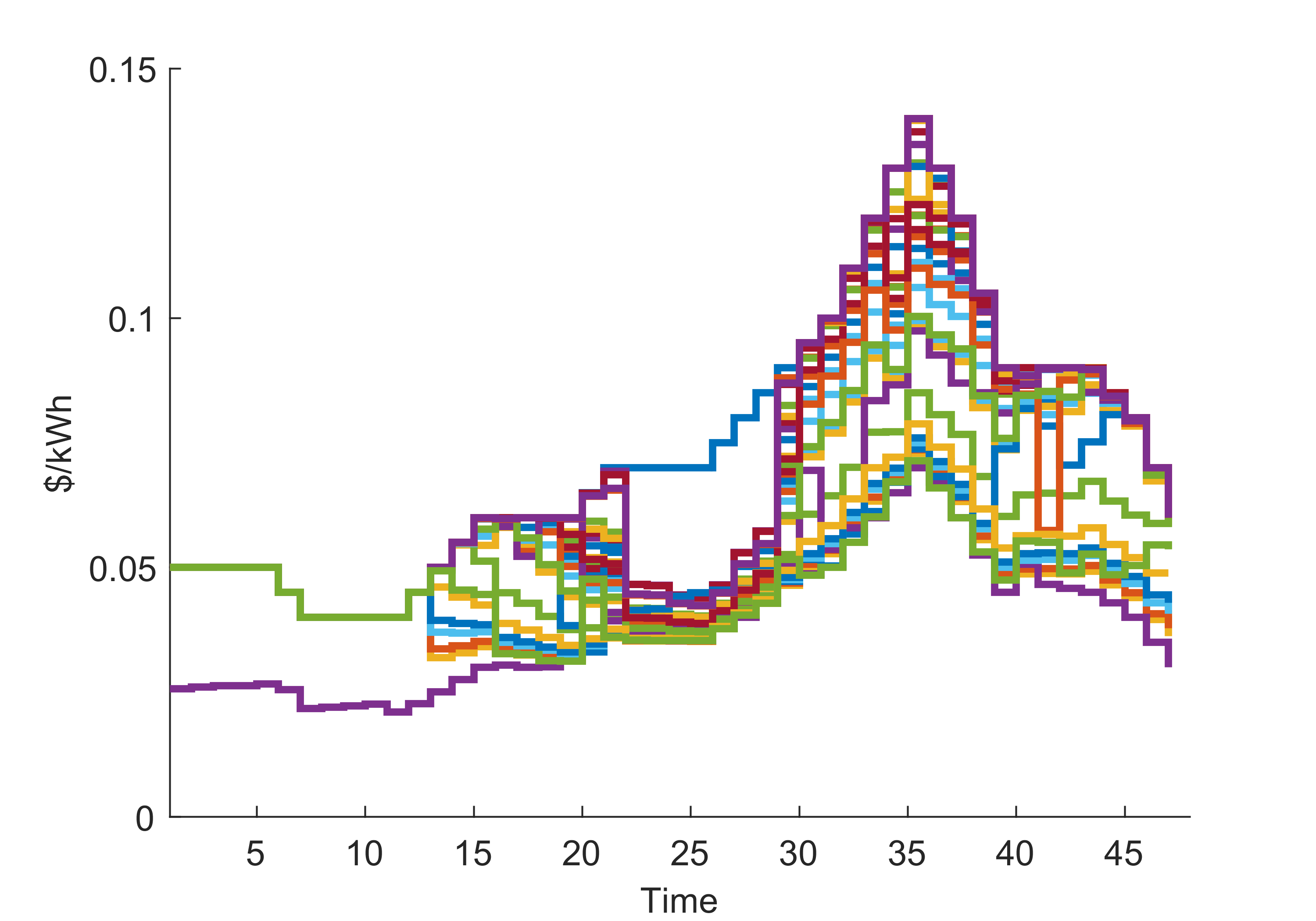}
    \caption{Transaction prices for each agent before benefit allocation}
    \label{fig: price_before_benalloc}
\end{figure}

\begin{figure}[] 
    \centering
    \includegraphics[width=0.45\textwidth]{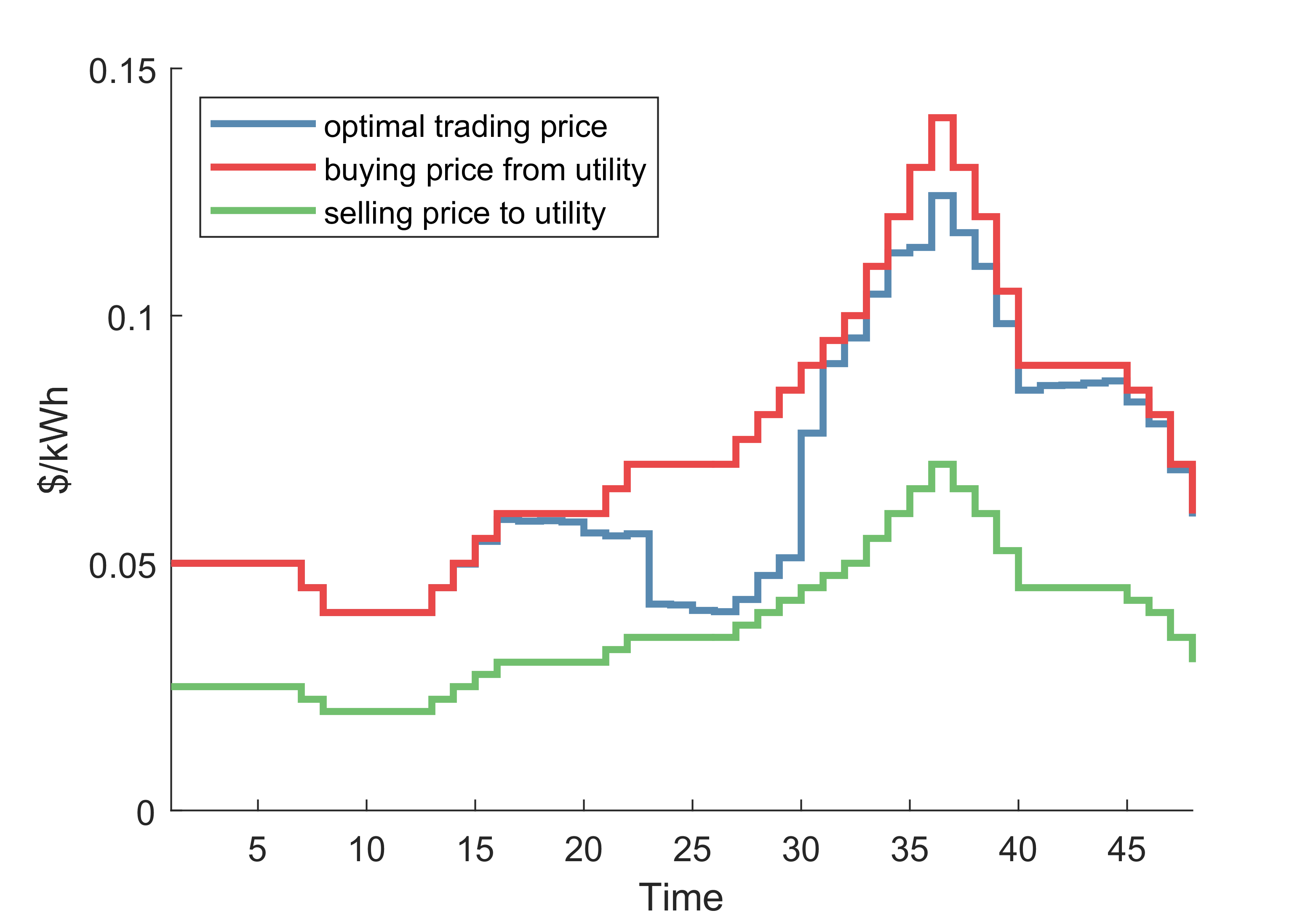}
    \caption{Comparison of optimal trading price $\pi^*$ after benefit allocation with utility buying price $u_b$ and selling price $u_s$}
    \label{fig: price_optimal_utility}
\end{figure}

The initial transaction price for each agent is shown in Fig. \ref{fig: price_before_benalloc} wherein we can see that the price varies for each agent. These are then used to calculate initial payments/incentives for imported/exported energy. With the benefit allocation algorithm, the optimal price $\pi^*$ is solved and this is uniform for all agents. This means that there exists an optimal price $\pi^*$ that upholds fairness in the system. We can also notice from Fig. \ref{fig: price_optimal_utility} that the optimal price solved is generally lower than the buying price from utility and higher than the selling price to utility, proving that trading is beneficial for the participants. There is significant difference between $\pi^*$ and $u_b$ in the middle of the day due to cheaper source of energy from excess PV. Notice that participants with excess PV still has better benefits in trading compared to transacting to the utility  since $\pi^*$ is still greater than $u_s$. Moreover, during at night, participants with DG and battery storage dispatch their DERs to benefit from profitable pricing while consumers also take the benefit from lower price of electricity compared to buying energy from the utility.

\begin{figure}[] 
    \centering
    \includegraphics[width=0.45\textwidth]{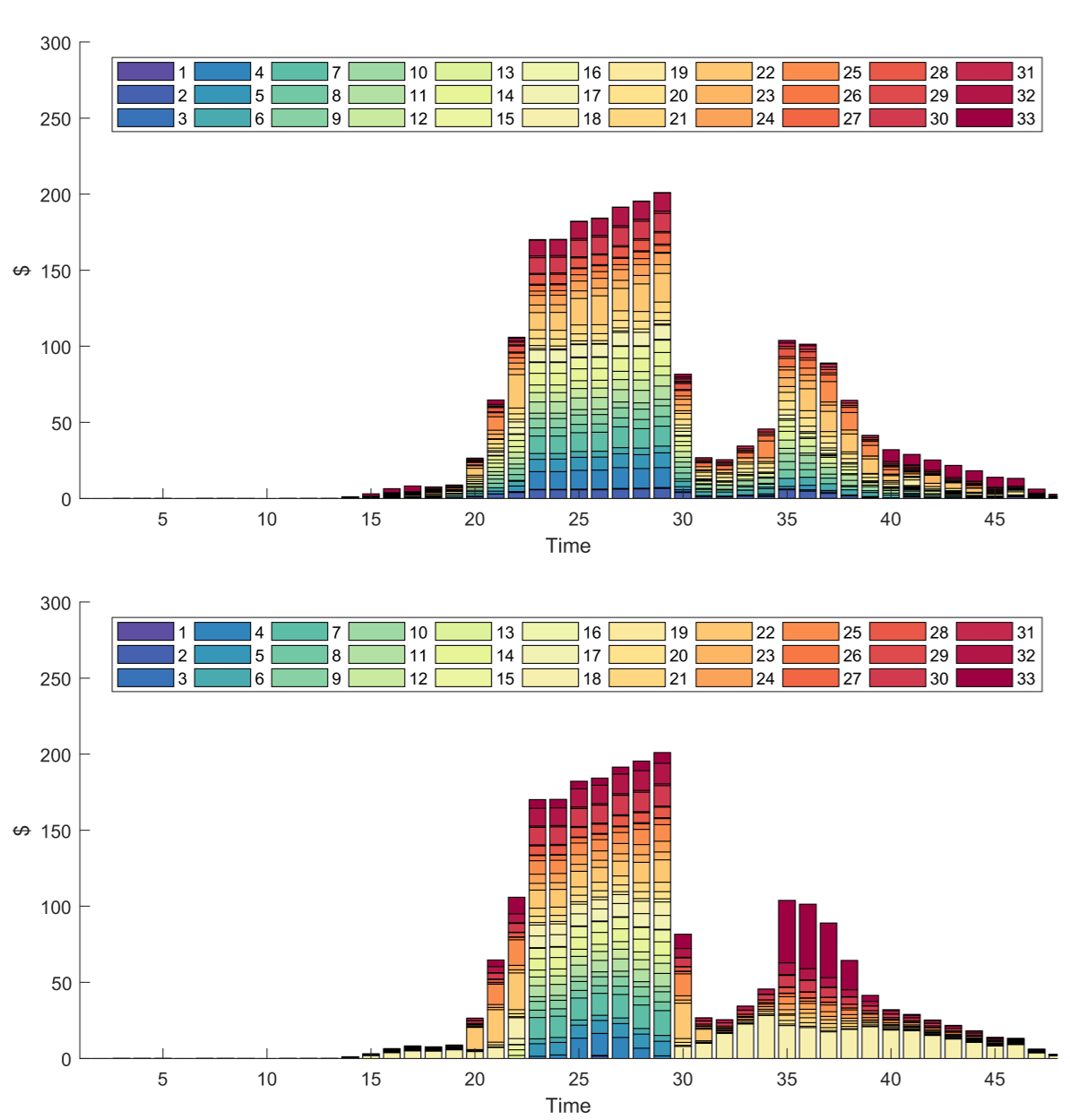}
    \caption{Profit before (top) and after (bottom) benefit allocation}
    \label{fig: profit_benefit_allocation}
\end{figure}

\subsubsection{Comparison of Profits}
For this part,we compare the profits before and after the last control stage. From Fig. \ref{fig: profit_benefit_allocation}, before benefit allocation, almost all buses share equal profits regardless of their contribution in the trading process. There is no clear distinction that buses with DERs gain more profits or being incentivized fairly in exporting excess energy from DER resources. We take for example at time $t=23$ where microgrid buses 18 and and 25 exported more than 1000 kWh each but only gained profits of  \$0.92 and \$2.80, respectively, as compared to consumer bus 17 that gained a profit of \$7.85 for importing 276 kWh. This type of scenario may discourage participants to participate in trading especially those with DERs. To address this issue, the benefit allocation algorithm is applied in the last control stage to ensure that trading participants receive appropriate benefits and incentives with respect to their contribution in the trading process. The second graph of Fig. \ref{fig: profit_benefit_allocation} particularly at time window of $t=39$ to $t=48$ shows a distinction of profit shares for microgrid buses through proper allocation of incentives with respect to the amount of energy exported during trading.

\subsubsection{Fairness}
\begin{figure*}[] 
    \centering
    \includegraphics[width=0.85\textwidth]{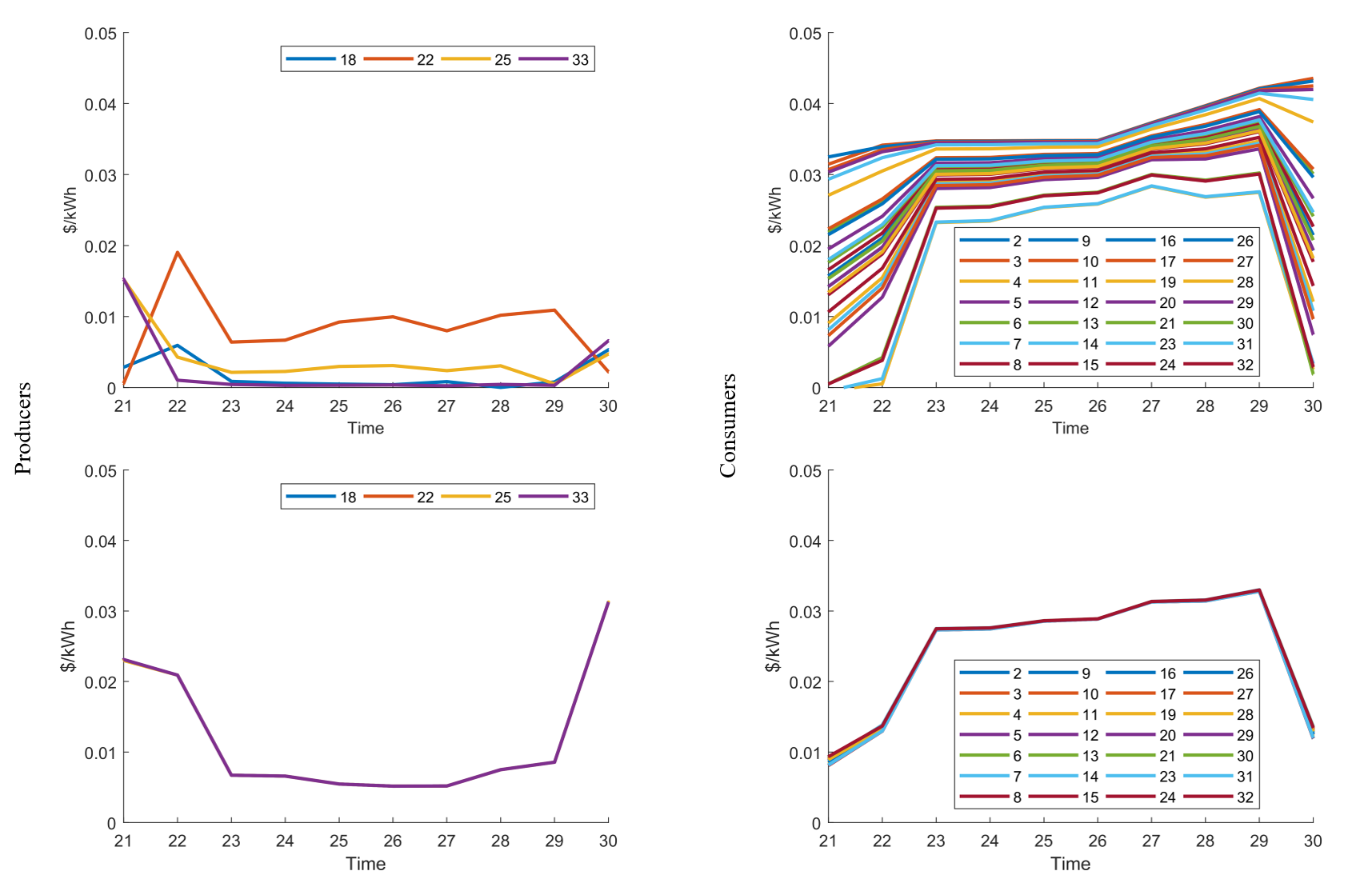}
    \caption{Profit per unit of energy transacted before (top) and after (bottom) benefit allocation for the set of producers and consumers}
    \label{fig:fairness}
\end{figure*}

In this part, we take a look at the effect of benefit allocation algorithm in the fairness of trading among participants in the system. With the algorithm, the profits of the set of consumers/producers with respect to the amount of energy imported/exported together with the incurred power losses, must almost be equal at each time step to achieve fairness in the system. Shown in Fig. \ref{fig:fairness} are the plots of profit per unit energy transacted for the set of producers and consumers before and after benefit allocation algorithm is applied. We consider the data at time window of $t=21$ to $t=30$ where all buses with DERs exported power simultaneously. This is for easier visualization and to have a clear distinction between producers and consumers. Notice that before benefit allocation algorithm is applied, the profits per unit energy transacted are not uniform for producers and consumers. But by using the algorithm, profits per unit energy transacted are equal and uniform for producers and consumers. It can be observed that the producers have less profit per unit energy transacted in the middle of the day since the optimal price calculated is low due to the surplus amount of PV. On the other hand, this will be beneficial for consumers to procure cheaper electricity, hence the increase in profit at the time window. The profits for producers start to increase at late afternoon to night where utility price start to peak and batteries and DGs are dispatched to provide cheaper electricity for consumers while giving more incentives to producers. With this, owners with DER assets are assured that they are incentivized fairly and more entities may be encouraged to install DERs and participate in trading using the proposed framework. This in turn will be more beneficial for the system to increase resilience and fully achieve energy independence.



\section{Conclusion and Future Work}
In this paper, we designed a three-stage energy trading framework for distribution systems that considers full AC network and power flow models and constraints addressing the gaps on existing transactive energy frameworks where network and power flow models used were relaxed and linearized . In addition, the concept of fairness in energy trading in a full AC network model was also investigated which has not been done in previous studies. A benefit allocation algorithm was integrated to ensure equal profits per unit energy imported or exported for the set of consumers and producers, therefore upholding fairness in the system. In the proposed framework, the cost without trading is calculated in the first stage and will then be used in solving the profit maximization problem in the second stage where the optimal network settings together with initial transaction payments and incentives are determined. At the last stage, optimal transaction price is calculated to determine fair payments and incentives for the participants based on their trading contribution in the system. Simulation results show that using the proposed framework led to reduction of overall energy cost for participants and balanced cash flow for the whole system. Moreover, the optimal trading price solved makes the energy trading beneficial for all participants since the system ensures that the trading price is higher than the selling price to utility but lower than the buying price from utility. The physical and economic aspects of energy trading has been investigated in this study and using the proposed framework, network constraints are satisfied and the concept of fairness has been guaranteed.

For this study, deterministic approach was used and uncertainties were not considered. Authors are now working on dealing with stochastic nature of system parameters such as load demand, renewable generation and utility price in energy trading.


%





\ifCLASSOPTIONcaptionsoff
  \newpage
\fi



%



\bibliographystyle{ieeetr}
\bibliography{main}

%








\end{document}